\documentclass[10pt,preprint]{aastex}

\def\ho{\ \mathrm{km}\ \mathrm{s}^{-1}\ \mathrm{Mpc}^{-1}}
\def\ni{\noindent}

\shorttitle{$H_0$ from linear \& non-linear PL relation}
\shortauthors{Ngeow \& Kanbur}

\begin{document}

\title{The Hubble Constant from Type Ia Supernova Calibrated with the Linear and Non-Linear Cepheid Period-Luminosity Relations}

\author{C. Ngeow}
\affil{Department of Astronomy, University of Illinois, Urbana, IL 61801}

\and

\author{S. M. Kanbur}
\affil{State University of New York at Oswego, Oswego, NY 13126}

\begin{abstract}

It is well-known that the peak brightness of the Type Ia supernovae calibrated with Cepheid distances can be used to determine the Hubble constant. The Cepheid distances to host galaxies of the calibrating supernovae are usually obtained using the period-luminosity (PL) relation derived from Large Magellanic Cloud (LMC) Cepheids. However recent empirical studies provide evidence that the LMC PL relation is not linear. In this Letter we determine the Hubble constant using both the linear and non-linear LMC Cepheid PL relations as calibrating relations to four galaxies that hosted Type Ia supernovae. Our results suggest that the obtained values of the Hubble constant are similar. However a typical error of $\sim0.03$mag. has to be added (in quadrature) to the systematic error for the Hubble constant when the linear LMC PL relation is used, assuming that the LMC PL relation is indeed non-linear. This is important in minimizing the total error of the Hubble constant in the era of precision cosmology. The Hubble constants calibrated from the linear and non-linear LMC PL relation are $H_0 = 74.92\pm2.28\mathrm{(random)}\pm5.06\mathrm{(systematic)} \ho$ and $H_0 = 74.37\pm2.27\mathrm{(random)}\pm4.92\mathrm{(systematic)} \ho$, respectively. Hubble constants calculated using the Galactic PL relations are also briefly discussed and presented in the last section of this Letter.

\end{abstract}


\section{Introduction}

Due to their intrinsic brightness at maximum, it is well-known that the Type Ia supernovae (SNIa) can be used to obtain the Hubble constant ($H_0$). Furthermore, the peak brightness of the Type Ia SN is regarded as a standard candle after correcting for their light curve shape \citep[for example: $\Delta m_{15}(B)$, $s$-factor, $MLCS$ and $CMAGIC$ in][respectively]{phi93,per97,rie96,wan03}. Nevertheless, the peak brightness of the supernovae after the light curve shape (and extinction) corrections need to be calibrated with nearby samples before they can be applied to derive the Hubble constant.

A common way to calibrate the peak brightness of SNIa is by using the period-luminosity (PL) relation from Cepheid variables in the galaxies that host the supernova \citep[see, e.g.,][and the reference therein]{gib00,sah01,rie05}. The current most widely used PL relation is derived from Large Magellanic Cloud (LMC) Cepheids. For a long time, the LMC PL relation has been regarded as linear in $\log(P)$, where $P$ (in days) is the pulsation period of the Cepheids. However, recent empirical studies have implied that the LMC PL relation is not linear: the LMC PL relation can be broken into two relations, for short ($\log[P]<1.0$) and long period LMC Cepheids, respectively \citep{tam02,kan04,san04,nge05,kan06}. Various rigorous statistical tests have been performed and the results strongly suggest that this non-linearity is real and not due to other factors such as extinction errors or a small number of long period Cepheids\footnote{Furthermore, the LMC period-color (PC) relation is also non-linear.}. Therefore, it is of great interest to examine how the non-linear LMC PL relation affects the calibration of the Hubble constant. 

Currently there are two studies that deal with this problem: \citet[][hereafter NK05]{nge05w} and \citet[][hereafter R05]{rie05}. In the former study, the authors examined the linearity of the LMC Wesenheit function, a linear combination of PL and PC relations. The Wesenheit function is frequently applied to derive Cepheids distances because it is reddening free \citep[see. e.g.][]{fre01,sah01,kan03,leo03}. NK05 found that the Wesenheit function for the LMC Cepheids is linear because the non-linearity of the LMC PL and PC relations almost cancel out. NK05 also suggested that the effect of a non-linear LMC PL relation in distance scale applications is minimal (at $\sim0.03$mag. level).  However, the authors did not go a step further to compare the Hubble constant calibrated from the linear and non-linear LMC PL relation and show that this is indeed the case. In contrast, R05 used the long period part of the non-linear LMC PL relation (their ``OGLE$+10$'' PL relation) to calibrate the peak brightness of the SNIa and hence derive the Hubble constant. But their study lacks a detailed comparison of the effect of linear vs. non-linear PL relation in distance scale applications. Therefore the main purpose of this Letter is to bridge the gap between these two studies by comparing the Hubble constants calibrated from the linear and non-linear LMC PL relation.

\section{Data \& Analysis}

Following the prescription given in R05, the Hubble constant from SNIa can be obtained with the following equation:

\begin{eqnarray}
\log H_0 = 0.2M^0_{\lambda}(t_{max}) + 5 + a_{\lambda},
\end{eqnarray}

\ni where $M^0_{\lambda}(t_{max})$ is the extinction corrected absolute magnitude (in bandpass $\lambda$) at peak brightness, and $a_{\lambda}\equiv \log (cz)-0.2m^0_{\lambda}(t_{max})$ is the distance-scale free intercept parameter determined from the ``distant'' supernovae that are located well within the Hubble flow \citep[see, e.g.,][]{jha99,rei05}. Here, we adopt the same value of $a_V=0.697\pm0.005$ as in R05. This value is determined from 38 SNIa in the Gold Sample of \citet{rie04}. Therefore, once the value of $M^0_V(t_{max})$ is calibrated with a Cepheid distance, the Hubble constant can be obtained in a straight forward manner. 

R05 also listed 4 ``ideal'' SNIa for the purpose of calibrating the $M^0_V(t_{max})$: they are SN 1994ae (in NGC 3370), SN 1998aq (in NGC 3982), SN 1981B (in NGC 4536) and SN 1990N (in NGC 4639). Since $m^0_V(t_{max})-M^0_V(t_{max})=\mu_{Ceph,0}$, where $m^0_V(t_{max})$ has been corrected for extinction and light curve shape of individual supernova, and because we are only interested in the changes of $\mu_{Ceph,0}$, this equation can be re-written as $M^0_V(t_{max})+\mu_{Ceph,0}=m^0_V(t_{max})=\mathrm{a\ constant}$ (with the constant term uniquely determined from the observations of individual supernova). Table 13 of R05 has gives the values of $M^0_V(t_{max})$ and $\mu_{Ceph,0}$ for these four SNIa, hence the change of $M^0_V(t_{max})$ due to the recalibration of $\mu_{Ceph,0}$ is just $M^0_V(new)=M^0_V(R05)+\mu_{Ceph,0}(R05)-\mu_{Ceph,0}(new)$. 

To obtain $\mu_{Ceph,0}(new)$, we use four sets of LMC PL relations as given in NK05\footnote{Except for KNB05 that we have updated the PL relations in \citet{kan06}, and referred this updated version as KN06 in this Letter.}. Each set of PL relations contain both the linear and non-linear (i.e., the long period PL relation) version of the LMC PL relation. The Cepheid data for these four galaxies are adopted from the following sources: NGC 3370 from R05; NGC 3982 from \citet{ste01}; and NGC 4536 \& NGC 4639 from \citet{gib00}. As in \citet{fre01}, we apply a period cut to Cepheids in NGC 3982, NGC 4536 \& NGC 4639 to avoid the incompleteness bias at the faint end of the Cepheid PL relation (there is no need to do this for NGC 3370, see R05). After fitting the PL relations to the data, we obtain the distance modulus via the Wesenheit function (see, e.g., the reference in NK05), $\mu_0=\mu_V-2.45(\mu_V-\mu_I)$. Metallicity corrections to $\mu_0$ were done in the same manner as in R05 (i.e., using the values in their table 8). Finally, the CTE (charge transfer efficiency) correction of $-0.07$mag. is applied to NGC 4536 \& NGC 4639 \citep{gib00,fre01}. No CTE correction is needed for NGC 3370 \& NGC 3982. Our results of the $\mu_{Ceph,0}(new)$, $M^0_V(new)$, $\log H_0$ from equation (1) and $H_0$ are summarized in Table \ref{tab1}.

\begin{deluxetable}{lccccccccc}
\tabletypesize{\scriptsize}
\tablecaption{Results of using the linear vs. non-linear LMC PL relation\tablenotemark{a}.\label{tab1}}
\tablewidth{0pt}
\tablehead{
\colhead{} & \colhead{} & \multicolumn{4}{c}{Linear LMC PL Relation} & \multicolumn{4}{c}{Non-Linear LMC PL Relation} \\
\colhead{NGC/SN} & \colhead{$N_{Ceph}$} 
& \colhead{$\mu_{Ceph,0}$} & \colhead{$M^0_V(t_{max})$} & \colhead{$\log H_0$}  & \colhead{$H_0$}
& \colhead{$\mu_{Ceph,0}$} & \colhead{$M^0_V(t_{max})$} & \colhead{$\log H_0$}  & \colhead{$H_0$}
}
\startdata
 & & \multicolumn{8}{c}{TR02 PL Relations} \\ 
NGC3370/SN1994ae & 64 & $32.193\pm0.033$ & -19.083 & 1.880 & 75.9 & $32.222\pm0.034$ & -19.112 & 1.875 & 75.0 \\
NGC3982/SN1998aq & 29 & $31.634\pm0.064$ & -19.164 & 1.864 & 73.1 & $31.658\pm0.064$ & -19.188 & 1.859 & 72.3 \\
NGC4536/SN1981B  & 35 & $30.810\pm0.043$ & -19.160 & 1.865 & 73.3 & $30.834\pm0.043$ & -19.184 & 1.860 & 72.4 \\
NGC4639/SN1990N  & 14 & $31.665\pm0.077$ & -19.025 & 1.892 & 78.0 & $31.699\pm0.076$ & -19.059 & 1.885 & 76.7 \\
 & & \multicolumn{8}{c}{STR04 PL Relations} \\ 
NGC3370/SN1994ae & 64 & $32.171\pm0.034$ & -19.061 & 1.885 & 76.7 & $32.149\pm0.033$ & -19.039 & 1.889 & 77.4 \\
NGC3982/SN1998aq & 29 & $31.608\pm0.064$ & -19.138 & 1.869 & 74.0 & $31.594\pm0.063$ & -19.124 & 1.872 & 74.5 \\
NGC4536/SN1981B  & 35 & $30.784\pm0.043$ & -19.134 & 1.870 & 74.1 & $30.770\pm0.044$ & -19.120 & 1.873 & 74.6 \\
NGC4639/SN1990N  & 14 & $31.648\pm0.076$ & -19.008 & 1.895 & 78.5 & $31.615\pm0.077$ & -18.975 & 1.902 & 79.8 \\
 & & \multicolumn{8}{c}{KN04 PL Relations} \\ 
NGC3370/SN1994ae & 64 & $32.189\pm0.033$ & -19.079 & 1.881 & 76.0 & $32.180\pm0.034$ & -19.070 & 1.883 & 76.4 \\
NGC3982/SN1998aq & 29 & $31.627\pm0.064$ & -19.157 & 1.866 & 73.5 & $31.618\pm0.064$ & -19.148 & 1.867 & 73.6 \\
NGC4536/SN1981B  & 35 & $30.803\pm0.043$ & -19.153 & 1.866 & 73.5 & $30.794\pm0.043$ & -19.144 & 1.868 & 73.8 \\
NGC4639/SN1990N  & 14 & $31.663\pm0.076$ & -19.023 & 1.892 & 78.0 & $31.656\pm0.076$ & -19.016 & 1.894 & 78.3 \\
 & & \multicolumn{8}{c}{KN06 PL Relations} \\ 
NGC3370/SN1994ae & 64 & $32.195\pm0.034$ & -19.085 & 1.880 & 75.9 & $32.213\pm0.034$ & -19.103 & 1.876 & 75.2 \\
NGC3982/SN1998aq & 29 & $31.633\pm0.064$ & -19.163 & 1.864 & 73.1 & $31.643\pm0.065$ & -19.173 & 1.862 & 72.8 \\
NGC4536/SN1981B  & 35 & $30.808\pm0.043$ & -19.158 & 1.865 & 73.3 & $30.819\pm0.042$ & -19.169 & 1.863 & 72.9 \\
NGC4639/SN1990N  & 14 & $31.670\pm0.076$ & -19.030 & 1.891 & 77.8 & $31.697\pm0.076$ & -19.057 & 1.886 & 76.9 \\
\enddata
\tablenotetext{a}{Errors in $\mu_0$ are random errors only. Unit of $H_0$ is in$\ho$. TR02 = \citet{tam02}; STR04 = \citet{san04}; KN04 = \citet{kan04}; KN06 = \citet[][this is the updated version of KNB05 in NK05]{kan06}.}
\end{deluxetable}

\begin{deluxetable}{lcccc}
\tabletypesize{\scriptsize}
\tablecaption{Error budget for the Hubble constant.\label{tab2}}
\tablewidth{0pt}
\tablehead{
\colhead{Source} &  \colhead{NGC3370/SN1994ae} & \colhead{NGC3982/SN1998aq} & \colhead{NGC4536/SN1981B} & \colhead{NGC4639/SN1990N}        
}
\startdata
 & \multicolumn{4}{c}{Random Error} \\ 
R1: Cepheid distance        & 0.034 & 0.065 & 0.044 & 0.077 \\
R2: SN light curve fit      & 0.12  & 0.12  & 0.12  & 0.12  \\
\tableline
Total $R$                   & 0.125 & 0.136 & 0.128 & 0.143 \\
\tableline
 & \multicolumn{4}{c}{Systematic Error} \\
S1: LMC distance                            & 0.10  & 0.10  & 0.10  & 0.10  \\
S2: Linear vs non-linear PL\tablenotemark{a}& 0.03 & 0.03 & 0.03 & 0.03  \\ 
S3: Metallicity correction                  & 0.03 & 0.03 & 0.05 & 0.08 \\
S4: Hubble flow($=5a_V$)                    & 0.025 & 0.025 & 0.025 & 0.025 \\
S5: $HST$ camera zero-point                 & 0.086 & 0.086 & 0.086 & 0.086 \\ 
\tableline
Total $S$ (linear PL)                       & 0.141 & 0.141 & 0.146 & 0.159 \\
Total $S$ (non-linear PL)                   & 0.138 & 0.138 & 0.143 & 0.156 \\
\enddata
\tablenotetext{a}{Only applicable when using the linear LMC PL relation.}
\end{deluxetable} 

The random errors that contribute to the Hubble constant include the random error in Cepheid distance modulus and the error from supernova light curve fit. Since the random errors in distance moduli, as given in Table \ref{tab1}, are almost identical to each other when using either the linear or non-linear PL relation (among the four sets of PL relations), we adopt a single value for the error for each galaxy. The random errors from supernova light curve fits are given in R05 with values of $0.12$mag. for each calibrator. The adopted values of the random errors for the four galaxies/calibrators are listed in Table \ref{tab2}. The systematic errors are discussed below, and they are summarized in Table \ref{tab2} as well. These systematic errors are:

\begin{enumerate}
\item Distance to the LMC: the PL relations given in NK05 are based on the LMC distance of $18.50\pm0.10$mag., as adopted by \citet{fre01}, hence we continue to adopt the conservative value of $0.10$mag. as the uncertainty in LMC distance.

\item Linear vs. non-linear PL relation: Since there is growing evidence that the LMC PL relation is non-linear, we assume that the non-linear LMC PL relation is the true underlying PL relation. Then there is additional $\sim 0.03$mag. systematic error for the derived distance modulus when using the linear version of the LMC PL relation (NK05)\footnote{This can be seen from, for example, Table \ref{tab1}. For a given set of PL relations, the distance moduli from the linear PL relation are systematically closer/further than the distance moduli, by $\sim0.01$mag. to $\sim0.03$mag., from the non-linear PL relation among the four calibrators.}. This error is not applicable to the distance modulus (or the Hubble constant) when using the non-linear PL relation.

\item Hubble flow: R05 determines the error of the Hubble flow from their Gold Sample is about $0.025$mag., which we adopted this value as well.

\item $HST$ camera zero-points: The zero-point uncertainties in $V$- and $I$-band for $HST$ cameras (both $ACS$ and $WFPC2$) are $0.03$mag., and the total uncertainty in Cepheid distance measurements is $\sqrt{(1.45\sigma_V)^2+(2.45\sigma_I)^2}=0.086$mag. for all four galaxies.

\item Metallicity correction: Uncertainties of the metallicity corrections are adopted from table 8 of R05. Note that this uncertainty is considered as a systematic error \citep{fre01,leo03} but not random error.

\end{enumerate}

The random errors for the four calibrators are used as the weights when calculating the weighted mean of the Hubble constant. This procedure also combines the random errors for individual calibrators into the overall random error on the Hubble constant. The overall systematic error in Hubble constant is adopted as the straight average of the systematic errors for the four calibrators. The results are shown in Table \ref{tab3} when using the four sets of linear and non-linear PL relation. 

\begin{deluxetable}{lccc}
\tabletypesize{\scriptsize}
\tablecaption{Hubble constant from various sets of LMC PL relations\tablenotemark{a}.\label{tab3}}
\tablewidth{0pt}
\tablehead{
\colhead{PL sets} &  \colhead{Linear LMC PL} & \colhead{Non-Linear LMC PL}  & \colhead{\% of variation} 
}
\startdata
TR02 & $74.96\pm2.28\pm5.07$ & $74.02\pm2.26\pm4.90$ & 1.3\% \\
STR04& $75.75\pm2.31\pm5.12$ & $76.47\pm2.33\pm5.06$ & 0.9\% \\
KN04 & $75.13\pm2.29\pm5.08$ & $75.44\pm2.30\pm4.99$ & 0.4\% \\
KN06 & $74.92\pm2.28\pm5.06$ & $74.37\pm2.27\pm4.92$ & 0.7\% \\
\enddata
\tablenotetext{a}{In unit of$\ho$, first and second errors are the random and systematic errors, respectively.}
\end{deluxetable} 

\section{Conclusion \& Discussion}

In this Letter we study the effect of linear vs. non-linear LMC PL relations in deriving the Hubble constant using SNIa. It can be seen from Table \ref{tab3} that the Hubble constants obtained from either the linear or the non-linear PL relation are consistent with each other (within the total error). The difference in the value of the Hubble constant obtained from the two methods is $\sim1.3$\% or less. Our results are consistent with the finding of NK05 that the non-linear PL relation has an minimal effect, at the $\sim0.03$mag. or at $\sim1$-$2$\% level, on distance scale studies for deriving the distance modulus and/or the Hubble constant. Assume that the true underlying LMC PL relation is indeed non-linear, with the latest evidence from \citet{nge06}, using the linear PL relation will change the distance modulus and hence the Hubble constant at the same $\sim1$-$2$\% level as compared to using the non-linear PL relation. Since the four sets of PL relations used in Table \ref{tab3} are not totally independent of each other (they share the same LMC Cepheid data from the Optical Gravitational Lensing Experiment), the Hubble constants in Table \ref{tab3} cannot be averaged. Here we adopt the Hubble constats from KN06 as our final results. Hence,

\begin{eqnarray}
H_0 \mathrm{(from\ linear\ PL)} & = & 74.92\pm2.28 \ \mathrm{(random)} \pm5.06 \ \mathrm{(systematic)}\ \ho , \nonumber \\
H_0 \mathrm{(from\ non-linear\ PL)} & = & 74.37\pm2.27 \ \mathrm{(random)} \pm4.92 \ \mathrm{(systematic)}\ \ho . \nonumber
\end{eqnarray}

\ni Note that our Hubble constants are consistent with the $H_0$ Key Project result \citep[][$72\pm8\ho$]{fre01} and the recent release of the 3-years WMAP result \citep[][$73.4^{+2.8}_{-3.8}\ho$]{spe06}.
 
It is well-known in the cosmology community that there is a degeneracy between $\Omega_{tot}$ (or $1-\Omega_k$) and $H_0$ (or $h$), as shown in \citet{teg04}. Hence in the era of precision cosmology, it is important to minimize errors on the estimation of Hubble constant from observation (to less than few percent level) to break this degeneracy. In this Letter we concentrate on studying the contribution to the error only from the form of the PL relation (linear vs. non-linear), and find that this could introduce an additional error at $\sim1$-$2$\% level to the total error. Table \ref{tab3} suggests that although the Hubble constants and the associated random errors are similar when using the linear vs. non-linear PL relation, the systematic errors are larger when using the linear PL relation if the LMC PL relation is indeed non-linear. Hence it is important to eliminate this additional error to improve the measurement of the Hubble constant to within few percent level. However there are other systematic errors that contribute to the total error, such as the uncertainty of the LMC distance ($0.10$mag.) which remains one of the largest systematic error in estimating the Hubble constant. Further refinement of all these systematic errors are clearly desired.
 
NK05 has described a way to derive Cepheid distances with non-linear PL relations when the target galaxy consist of both short and long period Cepheids. This can be done by using $\mu=\frac{1}{N_{short}}\sum \mu^{short}_i + \frac{1}{N_{long}}\sum \mu^{long}_j$, where $\mu^{short}$ and $\mu^{long}$ are the distance moduli for short and long period Cepheids respectively. Most of the $HST$ observed galaxies only contain the long period Cepheids, hence the long period part of the non-linear PL relation can be used. In addition to the distance scale studies, the existence of a non-linear PL relation in the LMC is very important for stellar pulsation and evolution studies: it is clearly crucial to investigate the underlying physics behind non-linear LMC PL relations \citep[see, e.g.,][]{kan04,kan06,nge05}.

We emphasize that the PL relations used in this Letter are the (linear and non-linear) LMC PL relations only. There are recent studies that suggest the Galactic Cepheids follow a different PL relation than then LMC Cepheids \citep{tam03,nge04,san04}, presumably due to the metallicity effects \citep[see, e.g.,][for the opposite point of view]{gie05}. \citet{kan03} compared the Cepheid distances to 25 $HST$ observed galaxies using both the (linear) LMC and the Galactic PL relations and found that there is, on average, a negligible ($\sim0.001$mag.) difference in the distance moduli when appropriate metallicity corrections are applied to the distance moduli from {\it both} of the LMC and the Galactic PL relations, which is first applied in \citet{kan03}. A similar result is also found in a recent paper by \citet{sah06}. This may suggest that the use of the Galactic and LMC PL relation could have minimal impact on the Hubble constant if the metallicity correction is applied (again, the contribution of using the {\it correct} vs. {\it incorrect} PL relations to the systematic error may be more important in reducing the total error on the Hubble constant). To see the effect of using the Galactic PL relation on the Hubble constant, we apply the same data and methodology as in Section 2, except that we replace the LMC PL relations with the Galactic PL relations. We adopt the recent Galactic PL relations from \citet{nge04} and the updated version in \citet{san04}.  The resulting Cepheid distances (with metallicity correction), the $M^0_V(new)$, $\log H_0$ and $H_0$ for the four calibrators are summarized in Table \ref{tab4}, with the layout similar to Table \ref{tab1}. The error budget for using the Galactic PL relation is very similar to the case of using the LMC PL relation (i.e., Table \ref{tab2}), except that there is no $0.03$mag. systematic error from the linear vs. non-linear PL relation. Furthermore, we adopt a conservative error of $0.10$mag. for the zero-point of the Galactic PL relation \citep{nge04,san04,sah06}, which includes various uncertainties from the distance measurements to individual Galactic Cepheids (for example, the open cluster fitting method, the infrared surface brightness method and the parallax measurements). The resulting Hubble constants are $H_0 = 70.91\pm2.16\mathrm{(random)}\pm4.69\mathrm{(systematic)} \ho$ and $H_0 = 69.60\pm2.12\mathrm{(random)}\pm4.60\mathrm{(systematic)} \ho$ with \citet{nge04} Galactic PL relations and \citet{san04} Galactic PL relations, respectively. The values of the Hubble constant from the Galactic PL relation are lower than those obtained from the LMC PL relations, however they are still consistent with each others within the $1\sigma$ of the total errors. 

\begin{deluxetable}{lccccccccc}
\tabletypesize{\scriptsize}
\tablecaption{Results of using the Galactic PL relation\tablenotemark{a}.\label{tab4}}
\tablewidth{0pt}
\tablehead{
\colhead{} & \colhead{} & \multicolumn{4}{c}{GAL-NK04 PL Relation} & \multicolumn{4}{c}{GAL-STR04 PL Relation} \\
\colhead{NGC/SN} & \colhead{$N_{Ceph}$} 
& \colhead{$\mu_{Ceph,0}$} & \colhead{$M^0_V(t_{max})$} & \colhead{$\log H_0$}  & \colhead{$H_0$}  
& \colhead{$\mu_{Ceph,0}$} & \colhead{$M^0_V(t_{max})$} & \colhead{$\log H_0$}  & \colhead{$H_0$}  
}
\startdata
NGC3370/SN1994ae & 64 & $32.324\pm0.035$ & -19.214 & 1.854 & 71.4 & $32.363\pm0.035$ & -19.253 & 1.846 & 70.1 \\
NGC3982/SN1998aq & 29 & $31.728\pm0.069$ & -19.258 & 1.845 & 70.0 & $31.768\pm0.069$ & -19.298 & 1.837 & 68.7 \\
NGC4536/SN1981B  & 35 & $30.902\pm0.039$ & -19.252 & 1.847 & 70.3 & $30.943\pm0.039$ & -19.293 & 1.838 & 68.9 \\
NGC4639/SN1990N  & 14 & $31.839\pm0.075$ & -19.199 & 1.857 & 71.9 & $31.877\pm0.075$ & -19.237 & 1.850 & 70.8 \\
\enddata
\tablenotetext{a}{Errors in $\mu_0$ are random errors only. Unit of $H_0$ is in$\ho$. GAL-NK04 = \citet{nge04}; GAL-STR04 = \citet{san04}.}
\end{deluxetable}

The apparent discrepancy between our results and the results presented in \citet{kan03} is due to the number of galaxies (25 vs. 4) available in both studies. \citet{kan03} found that if the average of $\log(P)$ from the Cepheids in a given galaxy is close to $\sim1.4$, then the difference in the distance moduli from the LMC and the Galactic PL relations will become negligible after the metallicity correction (see equation [5] \& [6] in \citealt{kan03}). This is also true for an ensemble of galaxies. The average $\log(P)$ for all 25 galaxies in \citet{kan03} is indeed $\sim 1.4$ (see their table 14), therefore an average of $\sim0.001$mag. difference in Cepheid distances is obtained in their paper. For each of the 4 calibrating galaxies in this Letter, the average $\log(P)$ is greater than $1.46$, this makes the Cepheid distances from the Galactic PL relation to be further. Hence the resulted Hubble constant obtained with the Galactic PL relations will be lower than the Hubble constant obtained with the LMC PL relations.

\acknowledgements
The authors would like to thank N. Suntzeff, R. Kennicutt and the anonymous referee for useful suggestions. SMK acknowledges the support from HST-AR-10673.04-A.

\end{document}